\documentclass{ws-ijmpd}

\begin{document}

\markboth{Araudo, Bosch-Ramon \& Romero}
{Jet-cloud interactions in AGNs}

%
\catchline{}{}{}{}{}
%

\title{High-energy emission from jet-cloud interactions in AGNs}

\author{Anabella T. Araudo}

\address{Instituto Argentino de Radioastronom\'ia (CCT La Plata - CONICET),
C.C.5, 1894 Villa Elisa,  Buenos Aires, Argentina\\
Facultad de Ciencias Astron\'omicas y Geof\'{\i}sicas,
Universidad Nacional de La Plata, Paseo del Bosque, 1900 La Plata,
Argentina \\
aaraudo@fcaglp.unlp.edu.ar}

\author{Valent{\'\i} Bosch-Ramon}
\address{Max Planck Institut f\"ur Kernphysik, Saupfercheckweg
1, Heidelberg 69117, Germany \\
Valenti.Bosch-Ramon@mpi-hd.mpg.de}

\author{Gustavo E. Romero}

\address{Instituto Argentino de Radioastronom\'ia (CCT La Plata - CONICET),
C.C.5, 1894 Villa Elisa,  Buenos Aires, Argentina\\
Facultad de Ciencias Astron\'omicas y Geof\'{\i}sicas,
Universidad Nacional de La Plata, Paseo del Bosque, 1900 La Plata,
Argentina \\
romero@fcaglp.unlp.edu.ar}

\maketitle


\begin{abstract}
Active galactic nuclei present continuum and line emission. 
The emission lines are originated by gas located close to the
central super-massive black hole. Some of these lines are broad,
and would be produced in a small region called broad-line region.
This region could be formed by clouds surrounding the central 
black hole. In this work, we
study the interaction of such clouds with the base of the jets in 
active galactic 
nuclei, and we compute the produced high-energy emission. 
We focus on sources
with low luminosities in the inner jet regions, to avoid strong gamma-ray
absorption. 
We find that the resulting high-energy
radiation may be significant in Centaurus A. Also, this phenomenon 
might be behind the variable $\gamma$-ray emission detected in M87,
if very large dark clouds are present.
The detection of jet-cloud interactions in active galactic nuclei would
give information on the properties of the jet base and the
very central regions.
\end{abstract}


\section{Introduction}	

Active galactic nuclei (AGN) are extragalactic sources composed 
by a super-massive black hole (SMBH), 
an accretion disk and bipolar relativistic jets. Some
AGNs present continuum emission in
the whole electromagnetic spectrum, from radio to $\gamma$-rays. 
Besides the continuum radiation, AGNs also have optical and ultra-violet  
line emission.
Some of these lines are broad, emitted by gas moving at velocities
$v_{\rm g} > 1000$~km~s$^{-1}$ and located in a region close 
($d \sim 10^{17}$~cm) to the SMBH.
The structure of the matter in the broad line region (BLR) is not well 
known but some models assume that the gas
could be clumpy.  
Dense clouds, confined by the hot external medium or by 
magnetic fields, would be ionized by 
photons from the accretion disk producing the emission lines 
broadened by the movement of the clouds around the SMBH. 
In the particular case of Faranoff Riley (FR) I galaxies, 
where the accretion disks have low luminosities, the  
photoionization of the clouds will be inefficient to produce lines and
the clouds might be dark.  

Centaurus A (Cen A) and M87 are the closest AGNs, located at distances of 
$\sim 4$ and $\sim 16$~Mpc, respectively. 
These AGNs are classified as FR I radio sources and in the case of Cen A
the nuclear region is obscured by a dense torus of gas and dust. 
Although the BLRs of Cen A and M87 have not been detected~\cite{Alex}, 
clouds  with similar characteristic to those detected in FR II AGNs may 
surround the SMBH~\cite{wang}.

We are interested in the high-energy emission produced by
the interaction of 
these possibly dark clouds with the jets of the AGN. 
We focus here on Cen A and M87,
since their moderate accretion rate would imply reduced photon-photon 
opacities in the interaction region, allowing $\gamma$-rays to escape.
 Assuming standard parameters for the clouds and jets, we study
the main physical processes that take place as a consequence of the 
interaction, and calculate the expected high-energy emission.

\section{Jet-cloud interaction}

We consider that clouds with density $n_{\rm c} = 10^{10}$~cm$^{-3}$ 
and size $R_{\rm c} = 10^{13}$~cm are surrounding the SMBH 
and one of these clouds penetrates into one of the relativistic jets.  
We assume that the jet has a Lorentz factor $\Gamma = 10$ 
(i.e. a bulk velocity $v_{\rm j} \sim c$), a radius $R_{\rm j} = 0.1z$ 
($z$ is the distance to the black hole), 
and a kinetic luminosity $L_{\rm j} = 10^{44}$~erg~s$^{-1}$.

The penetration time of the cloud into the jet is determined by 
$t_{\rm c} \sim 2 R_{\rm c}/v_{\rm c} = 2\times10^4$~s, 
where $v_{\rm c} = 10^9$~cm~s$^{-1}$ is the cloud velocity.
As a consequence of the interaction of the jet material with the cloud, 
two shocks form.
One of these shocks propagates back in the reference frame of the jet 
with a velocity $v_{\rm bs} \sim v_{\rm j}$, producing a bow shock.
This bow shock reaches the steady state configuration  in a time 
$t_{\rm bs} \sim x_{\rm bs}/v_{\rm bs}$, where the stagnation point is 
taken at a distance $x_{\rm bs} \sim 0.3 R_{\rm c}$ from the cloud
(this value is obtained considering particle flux conservation and 
assuming an escape velocity equal to the downstream sound speed).  
On the other hand, a shock propagates inside the cloud at a velocity
$v_{\rm sc} \sim v_{\rm j} (\Gamma -1)/\chi $, 
where $\chi \equiv n_{\rm c}/n_{\rm j}$ and 
$n_{\rm j} = L_{\rm j}/(\pi R_{\rm j}^2(\Gamma -1)m_pc^2 v_{\rm j})$ 
is the jet density
in the shock reference frame. In a time 
$t_{\rm cc} \sim 2R_{\rm c}/v_{\rm sc}$ the whole cloud is shocked. 

The permanence of the cloud into the jet is determined by the passage time of 
the cloud in the jet, defined by $t_{\rm j} \sim 2R_{\rm j}/v_{\rm c}$. 
However, 
the cloud is accelerated by the jet, starting to move with the
outflow. The acceleration of
the cloud is $g \sim v_{\rm j}^2(\Gamma -1)/(\chi R_{\rm c})$.
The acceleration timescale can be estimated from
$t_{\rm g} \sim \sqrt{R_{\rm c}/g} \sim t_{\rm cc}$, which is the time 
required to accelerate the cloud up to $v_{\rm sc}$ in the jet direction,
hence one can assume that the bow shock will remain strong only during 
several times $t_{\rm cc}$.   
As a consequence of the pressure exerted by the jet onto the cloud, 
Rayleigh-Taylor (RT) instabilities can develop in the interface. 
In addition, Kelvin-Helmholtz (KH) instabilities can grow as a 
result of the high relative velocity between the jet
shocked material and the cloud.
In a first order approach, we obtain that RT and KH instabilities
grow sufficiently to destroy the cloud on a timescale 
$t_{\rm RT} \sim t_{\rm KH} \sim t_{\rm cc}$. In order to estimate the
lifetime of the cloud into the jet, we compare $t_{\rm j}$ and 
$t_{\rm cc}$. 
We can parametrize $t_{\rm cc} = \zeta t_{\rm c}$,
$\zeta > 1$, obtaining $v_{\rm sc} = v_{\rm c}/\zeta$.
Considering that  $v_{\rm sc}$ depends on $z$ through the parameter
$\chi$, we can obtain the interaction height:  
$z_{\rm int} = \zeta\, 2.5\times10^{15}$~cm.
Adopting $\zeta = 2$ results
$t_{\rm j} \sim 10^7$~s and $t_{\rm cc} \sim 2\times10^4$~s, i.e. the
cloud will be destroyed by the jet before escaping or approaching
the jet velocity (with the subsequent weakening of the shock). 

\begin{figure}[]
\centerline{\psfig{file=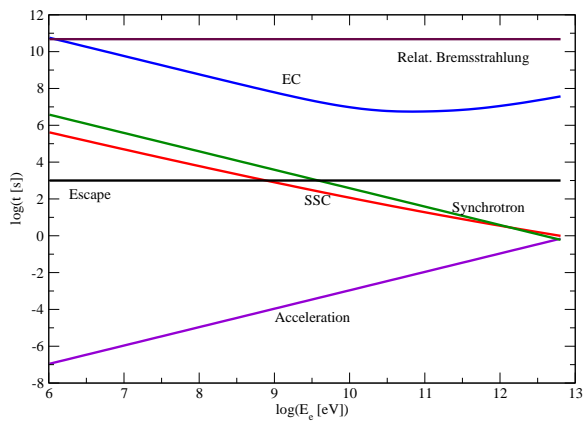,width=7cm, angle=270}}
\vspace*{8pt}
\caption{Acceleration and cooling timescales for relativistic electrons 
accelerated in the bow shock of the jet-cloud interaction.\label{f1}}
\end{figure}

\section{Particle acceleration}

In this work we consider only the particle acceleration in the 
bow shock, neglecting the contribution from the shock in the cloud. 
We assume that particles are accelerated up to 
relativistic energies,
being injected in the downstream region of the bow shock following a 
distribution
$Q_{e,p} \propto E_{e,p}^{-2.2}$ ($e$ and $p$ for electrons and protons, 
respectively).
Assuming that a $20\%$ of the jet luminosity that reaches the cloud goes 
to accelerate particles, the luminosity of these will be    
$L_{\rm nt} \sim 0.2(R_{\rm c}/R_{\rm j})^2L_{\rm j}$. 
Considering that the magnetic energy density in the bow-shock region is 
the 10\% of the non-thermal one, we obtain a magnetic field $B \sim 10$~G
in that region.

The main radiative losses that affect the evolution of
$Q_{e}$ are synchrotron radiation and synchrotron self-Compton (SSC) 
scattering. On the other hand, considering 
ultra-violet seed photons  with a luminosity 
$\sim 10^{42}$~erg~s$^{-1}$ (a larger value is unlikely
in the nuclear region of FR I galaxies) the external Compton 
(EC) cooling is not relevant, as shown in Fig.~\ref{f1}. 
Relativistic Bremsstrahlung losses produced by the interaction with the 
jet matter in the bow-shock downstream region is not important either. 
In addition to the radiative losses, electrons escape from the 
emitter on a time $t_{\rm esc} \sim 3 R_{\rm c}/c$.
As  we can see from  Fig.~\ref{f1}, the maximum energy is 
determined by synchrotron losses, given $E_e^{\rm max} \sim 1$~TeV. 
To obtain the distribution $N_e$ of relativistic electrons,
we solve the kinetic equation (see Ref.~\refcite{ginz})  
but taking into account not only the losses mentioned above but also the  
synchrotron photon field as SSC target, for which we need to use an 
iterative approach due
the non-linear nature of the problem to solve. We obtain 
that the steady state is reached on a timescale $\ll t_{\rm cc}$.
The energy distribution of electrons $N_e(E_e)$ has a break at energy 
$E_{\rm b} \sim 1$~GeV due to electron escape (see Fig.~\ref{f1}).
    
In the case of protons, these particles can lose energy via $pp$ interactions 
in the bow-shock region but the diffusion losses are more important, 
constraining the maximum energy to $E_{p}^{\rm max} \sim 3\times10^3$~TeV.
The most energetic protons, $E_p > 1$~TeV,
diffuse up to the cloud before escaping advected by the shocked 
material of the jet.

\section{High-energy emission}

In the bow-shock region, using standard formulae 
\cite{blu} and the electron energy distribution, we compute the 
synchrotron and SSC emission.
In the cloud, energetic protons that arrive from the bow shock 
are not confined and escape from the cloud on a time 
$t_{\rm cl} \sim R_{\rm c}/c$ before radiating a significant part of 
their energy. Considering that the distribution of protons in the cloud is 
$N_p \sim Q_p\,t_{\rm cl}$, we compute the $pp$ emission following
the formulae given in Ref.~\refcite{kelner}.

\begin{figure}[h!]
\centerline{\psfig{file=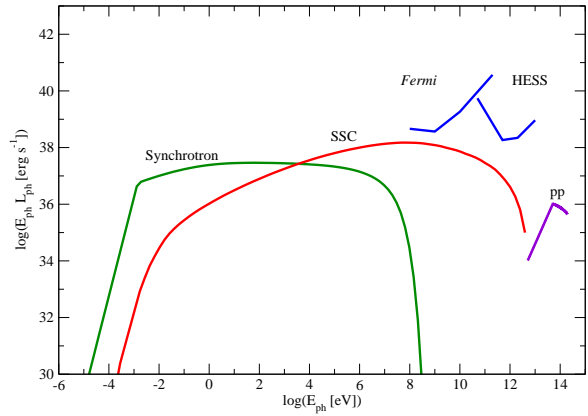,width=7cm, angle=270}}
\vspace*{8pt}
\caption{SED for the interaction of one cloud with the jet in
Cen A. Also we plotted the sensitivities of \emph{Fermi} and HESS
instruments, for 1 year and 50 hours of exposition, respectively.
\label{f2}}
\end{figure}

The synchrotron emission is self-absorbed at energies 
$E_{\rm ph} < 10^{-4}$~eV, but at higher energies 
auto-absorption and $\gamma\gamma$ absorption (see Ref.~\refcite{franck}) 
are  negligible in the region of interest. 
The achieved luminosity at energies $\sim 0.1-10$~GeV is 
$L_{\rm SSC} \sim 2\times10^{39}$~erg~s$^{-1}$, being  
slightly less than the sensitivity of HESS and \emph{Fermi} at the distance 
of Cen A as is shown in Fig.~\ref{f2}. 
Note that we show the result of the interaction of only one cloud with 
the jet, but many clouds could simultaneously interact with the jet.

\section{Discussion}

The total luminosity produced by jet-cloud interactions depends on the 
number of clouds inside the jet, each
one producing a spectral energy distribution (SED) with similar 
characteristics and luminosity levels to those shown in Fig. 2.  

\begin{figure}[]
\centerline{\psfig{file=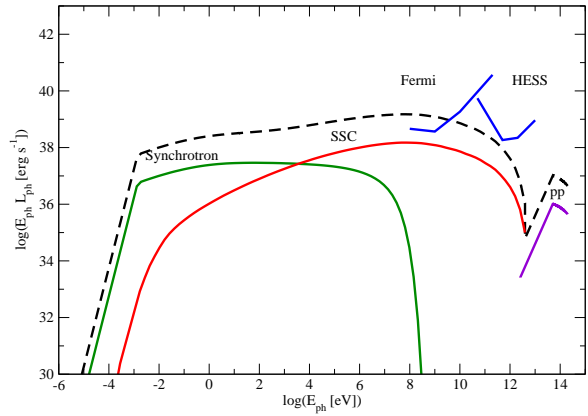,width=7cm, angle=270}}
\vspace*{8pt}
\caption{The same as in Fig.~\ref{f2} but for $\sim$~10 clouds.\label{f3}}
\end{figure}

The number of clouds inside the jet is $N_{\rm cj} = f V_{\rm j}/V_{\rm c}$,
where $f$ is the filling factor of 
dark clouds, and $V_{\rm j}$ and $V_{\rm c}$ are the jet and the cloud volume, 
respectively. Considering $V_{\rm j}$ up to $z \sim 10^{16}$~cm and
$f \sim 5\times10^{-6}$ (in FR II galaxies $f \sim 10^{-6}$), we obtain  
$N_{\rm cj} \sim 10$. In the whole sphere of size $\sim 10^{16}$~cm the 
number of clouds is $\sim 5\times10^3$ for the considered value of $f$.
The simultaneous interaction of $\sim$ 10 clouds 
 with the jet will produce more luminosity than the one produced by only
one interaction. If all the clouds have the same properties (i.e. $n_{\rm c}$,
$R_{\rm c}$ and $v_{\rm c}$) and are located at $z \sim z_{\rm int}$ in the jet, 
then the contribution of 10 clouds will produce a SED
with a similar appearence than that shown in Fig.~\ref{f2}, but with
a luminosity $\sim$ 10 times larger, being now detectable by HESS and 
\emph{Fermi} telescopes in the case of Cen A, as is shown in Fig.~\ref{f3}.
The emission detected by these instruments from Cen~A is 
larger than the luminosity predicted by our model.
However, if clouds are larger than $10^{13}$~cm, the luminosity level
detected by  HESS\cite{hess_CenA} and \emph{Fermi}\cite{fermi_CenA} 
could be achieved.
A more detailed calculation of the emission produced by many clouds 
interacting with the jet at different $z$ will be presented in a future
work\cite{Araudo}.

In the case of M87, the jet base is expected to be at 
$z_0 \sim 50 R_{\rm Sch} \sim 4\times10^{16}$~cm. At such a height on the
jet, the jet-cloud interaction
will be inefficient producing high-energy emission due to the small
cloud to jet section ratio. 
In order to obtain a detectable luminosity, clouds with a radius 
$\sim 10^{14}$~cm would be
necessary. In the case of a very big cloud entering the jet close to
$z_0$ in M87, the interaction might produce the variable 
luminosity detected by HESS\cite{hess_m87}.

\section*{Acknowledgments}

The authors thank Luigi Costamante for fruitful 
discussions. A.T.A. and G.E.R. are supported by the grant PICT 2007-00848.
G.E.R. and V.B-R.  acknowledge support by DGI
of MEC under grant AYA2007-68034-C03-01.
V.B-R. wants to thank the Instituto Argentino de
Radioastronom\'ia, and the Facultad de Ciencias Astron\'omicas y
Geof\'isicas de la Universidad Nacional de La Plata, for their kind 
hospitality.


\end{document}